\documentclass[%
 aip,
 amsmath,amssymb,
preprint,%
]{revtex4-1}

\usepackage{graphicx}
\usepackage{dcolumn}
\usepackage{bm}

\usepackage{graphicx}
\usepackage{subcaption}
\usepackage{mwe}
\usepackage{caption}

\usepackage[utf8]{inputenc}
\usepackage[T1]{fontenc}
\usepackage{mathptmx}
\usepackage{xcolor}

\begin{document}


\title[Single-laser scheme for reaching strong field QED regime via direct laser acceleration ]{Single-laser scheme for reaching strong field QED regime via direct laser acceleration}

	\author{R. Babjak}
	\email[]{robert.babjak@tecnico.ulisboa.pt}

	\affiliation{GoLP/Instituto de Plasmas e Fusão Nuclear, Instituto Superior Técnico, Universidade de Lisboa, Lisbon, 1049-001, Portugal}
	\affiliation{Institute of Plasma Physics, Czech Academy of Sciences, U Slovanky 2525/1a, 182 00 Praha 8, Czechia}

	\author{M. Vranic}
	\affiliation{GoLP/Instituto de Plasmas e Fusão Nuclear, Instituto Superior Técnico, Universidade de Lisboa, Lisbon, 1049-001, Portugal}

\date{\today}

\begin{abstract}

We investigate a single-laser scheme for reaching the strong-field QED regime based on direct laser acceleration (DLA) of electrons followed by their head-on collision with the same laser pulse reflected from an overdense foil. In this configuration, electrons are first accelerated inside an underdense plasma by a relativistic laser pulse and subsequently interact with the reflected laser field, emitting high-energy photons via nonlinear Compton scattering which decay into electron–positron pairs through the nonlinear Breit–Wheeler process. Using analytical scalings supported by quasi-3D particle-in-cell simulations including QED effects, we demonstrate that a laser pulse with power as low as 2 PW is sufficient to reach the quantum regime characterized by $\chi_e > 1$. For higher powers, we observe a rapid nonlinear increase in the number of generated positrons, reaching more than 2 nC for a 10 PW laser pulse with energy of approximately 1.1 kJ. A semi-analytical model is employed to estimate the positron yield, showing good agreement with simulation results. We further study the influence of laser depletion and the positioning of the reflecting foil on the efficiency of pair production. The presented scheme provides an experimentally feasible platform for probing strong-field QED effects using currently available multi-petawatt laser systems.

\end{abstract}

\maketitle


\section{Introduction}

The progress in the development of high-power laser pulses and their increasing availability at many facilities \cite{cernaianu2025,webber2017,zou2015,yoon2021} enables laboratory studies of laser-driven electron–positron pair creation. One of the potential applications is the study of fundamental interactions in the strong-field quantum electrodynamics (SFQED) regime \cite{nikishov1967,dipiazza2012,gonoskov2022}. It is achieved when a lepton, in its rest frame, feels an electric field exceeding the Schwinger critical field \cite{schwinger1951}. 

An important aspect of this type of interaction is the energy loss of leptons in a strong field due to emission, described either by the quantum-continuous \cite{baier1998} or by a quantum stochastic model \cite{gonoskov2022} of radiation reaction. Verifying radiation reaction models is of fundamental importance, with implications in many astrophysical scenarios, namely gamma-ray burst formation \cite{sultana2013}, magnetic reconnection \cite{lyubarskii1996}, or plasmas in magnetospheres of compact objects \cite{philippov2018}.

Another motivation for studying the SFQED regime is the possibility of generating pair plasmas using strong lasers to study their kinetic properties in the laboratory \cite{kirk2009,ridgers2012}. Alongside the use of thick high-$Z$ targets facilitating the Bethe–Heitler process to create such plasmas, the use of all-optical schemes to generate such states of matter is expected to be a promising option \cite{chen2023}.

There are several channels available to create electron–positron pairs, for example via photon–photon interaction. If two high-energy photons interact ($\gamma+\gamma' \rightarrow e^- + e^+$), the process is called the linear Breit–Wheeler (LBW) process \cite{breit1934}. The product of the two photon energies needs to exceed 0.25 MeV$^2$. The very small cross section of the LBW process poses demanding requirements on the yield of the gamma-photon source. Even though it was experimentally observed at a heavy-ion collider \cite{adam2023}, laser-based setups have not been successful yet, despite several schemes being proposed \cite{he2021,he2021_2,sugimoto2023}.

Pair production can be also achieved via the interaction of energetic photons with strong laser field. The processes present are nonlinear inverse Compton scattering (NICS) and nonlinear Breit–Wheeler (NBW) pair creation, since their probability strongly increases when the SFQED regime is attained. The NICS process consists of an electron absorbing $n$ laser photons and emitting a high-energy $\gamma$ photon ($e^- + n\omega_{\rm{las}} \rightarrow e^- + \gamma$), whilst NBW corresponds to a $\gamma$ photon and $n$ laser photons being converted into an electron–positron pair ($n\omega_{\rm{las}} + \gamma \rightarrow e^- + e^+$). It was observed for the first time at Stanford Linear Accelerator Center (SLAC), when 46 GeV electron beam accelerated by the linear accelerator collided with the terawatt laser pulse \cite{burke1997, bamber1999}. Even though there are currently ongoing attempts to reproduce the experiment with better quality of electron and laser beam at DESY \cite{levy2022} and at SLAC, there are not many facilities that have simultaneously access to a laser pulse with relativistic intensity and a linear accelerator capable of producing multi-GeV electron beams. To overcome this obstacle, all-optical methods have been proposed.

One of the all-optical methods to probe strong-field QED in a head-on collision setup of an electron and laser beam is based on splitting a laser pulse into two beams. After splitting, one laser beam is directed into a gas cell to accelerate electrons via LWFA to energies of the order of several GeV. The second part of the laser beam is focused to achieve high intensity in a spot size of the order of a few microns and is propagated against the accelerated electron beam \cite{vranic2014}. In this setup, electrons colliding with the high-intensity laser beam emit X-rays and $\gamma$ photons due to Compton scattering \cite{powers2014,cole2018}. If the photon energy and laser field intensity are high enough, photons can decay into electron–positron pairs in the strong laser field due to the nonlinear Breit–Wheeler process \cite{mirzaie2024,turner2022,los2024}.

Recently, a single-laser approach to probe strong-field QED effects was proposed \cite{matheron2025,gerstmayr2025}. After the laser acceleration stage in gas, the laser pulse is reflected from an over-critically dense plasma mirror, which ensures a head-on collision between the laser and accelerated electrons. This scheme has previously been used to generate X-ray sources based on Compton scattering \cite{taphuoc2012,meir2024}. For efficient LWFA in low-density gas, a laser spot size of the order of tens of microns is commonly used, which enables stable propagation and acceleration \cite{lu2007}. However, a wide spot size resulting in low laser intensity is not favourable for reaching the QED regime. To enable substantial pair creation, the laser needs to be focused after the interaction stage, for which plasma lenses have been proposed as a solution.

\begin{figure}
    \centering
    \includegraphics[width=0.95\linewidth]{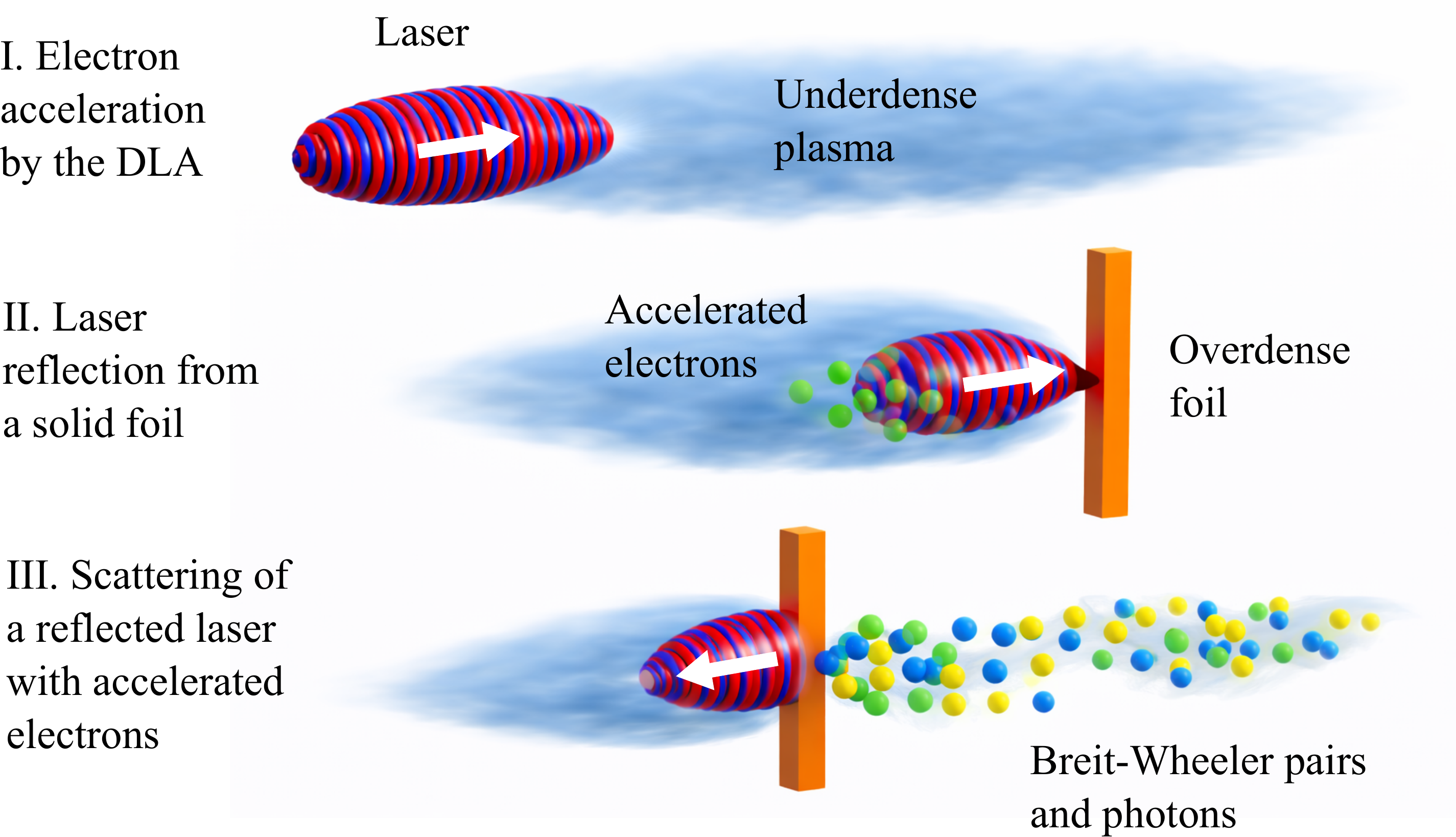}
    \caption{Illustration of the different stages studied: from single-laser DLA towards reaching the strong field QED regime. At first, electrons are accelerated inside the gas target by the relativistic laser pulse propagating for several millimetres. After the electrons are accelerated, an overdense foil is placed on the optical axes, to reflect the remainder of the laser pulse backwards. After the reflection, the electrons emit photons via non-linear Compton scattering. These photons then decay into electron-positron pairs via Breit-Wheeler process in the strong field environment provided by the laser pulse.  }
    \label{fig:compton_setup}
\end{figure}

In this paper, we examine the potential of DLA in the optimal regime aiming for the highest energy cut-off presented by Babjak et al. \cite{babjak2024}. The scheme is depicted in Fig. \ref{fig:compton_setup}. First, electrons are accelerated during the propagation of a laser pulse through gaseous plasma by the direct laser acceleration (DLA) mechanism \cite{pukhov1998,pukhov1999,khudik2016,arefiev2016,arefiev2016_2}, which has been extensively studied experimentally \cite{gahn1999,mangles2005,kneip2008,hussein2021,willingale2013,willingale2018,rosmej2021,gyrdymov2024,shou2023,rinderknecht2021}. Electrons can also be accelerated by continuously injected LWFA in scenarios when the laser pulse is too short for efficient DLA \cite{vais2024,horny2024}. After a certain distance, an overcritically dense target is placed in the laser propagation direction to reflect the laser. After reflection, the laser pulse interacts with electrons that were accelerated by DLA. First, photons are emitted due to nonlinear Compton scattering and consequently decay into electron–positron pairs via the nonlinear Breit–Wheeler process.

There are several advantages and disadvantages when compared to LWFA-based scheme. The acceleration in the DLA regime with multi-PW lasers is the most efficient with the moderate spot size of $\sim10$ microns. Using higher densities of 0.1 $n_c$ lowers requirements on the spot size even more, enabling the acceleration with spot size of $\sim5$ microns. This results in the acceleration with significantly higher intensity compared to the LWFA, which makes the laser focusing after the electron acceleration redundant, simplifying the experimental set up. Another advantage of the DLA-based approach is that electrons are being accelerated inside the laser pulse. This enables electrons to interact with the highest-intensity part of the laser pulse right after the reflection. In the LWFA-based scenario, electrons can be scattered away from the laser due to the ponderomotive force, wihtout reaching the highest intensity regions. Furthermore, high accelerated electron charge during the DLA process results in a high charge of generated positrons, simplifying their detection. Since the DLA mechanism is known for the ability to accelerate electron bunches with a high charge, they are better suited for reaching densities high enough to explore collective kinetic pair plasma effects. Downside of the mechanism is the intrinsic broad DLA spectrum, which is not well-suited for future high-fidelity tests of strong-field QED. Therefore we state that the approach presented in this work is not meant as a replacement of LWFA-based schemes, but rather as a complementary approach with its own range of applications.

The work is organized as follows. Section \ref{s:analitical_estimates} presents analytical estimates to predict electron properties during acceleration, along with scalings for the number of generated electron–positron pairs. Because the interaction occurs in a strongly nonlinear regime, potential discrepancies between the scalings and realistic scenarios are also discussed. Section \ref{s:depletion} presents quasi-3D PIC simulations of electron acceleration, where laser front depletion is demonstrated and compared with analytical predictions. Properties of electrons after the acceleration stage are discussed in Section \ref{s:particle_properties}, along with properties of positrons generated by the Breit–Wheeler process after laser–electron scattering. Since the number of generated positrons increases rapidly when the QED regime is reached, we estimate the minimal laser power requirements for reaching $\chi>1$ in Section \ref{s:qed}. In Section \ref{s:number_estimate}, the number of generated positrons is estimated semi-analytically and found to be in excellent agreement with simulation results. Lastly, Section \ref{s:mirror_position} discusses an optimization strategy for maximizing positron yield by placing the reflecting mirror at different positions. This strategy is developed using an additional set of PIC simulations.

\section{Analytical estimates for the number of generated positrons }\label{s:analitical_estimates}

The first stage of the scheme is direct laser acceleration of electrons in a gas. The first condition for electron acceleration by DLA to occur is that the laser pulse length needs to be longer than the length of the first bucket of a plasma wake, $\Lambda_{\mathrm{wake}}\simeq \sqrt{a_0}[2\pi/k_p]$, where $k_p=\omega_p/c$ \cite{shaw2017}. During the acceleration, electrons co-propagate with the laser pulse, performing resonant betatron oscillations while gaining energy directly from the laser field. While the transverse component of the electric field accelerates electrons, the magnetic field component of the laser bends their trajectories in the forward direction. For each combination of laser amplitude $a_0$ and background plasma density $n_e$, the maximum energy that electrons can achieve is

\begin{equation}\label{eq:gmax_resonant}
    \gamma_{\rm{max}}=2 \left( \frac{a_0}{\varepsilon_{cr}} \right)^{4/3} \left( \frac{n_e}{n_c} \right)^{-1/3}.
\end{equation}

To reach the energy limit achievable in the DLA regime $\gamma_{\rm{max}}$, electrons need to co-propagate with the laser at least for the distance

\begin{equation}\label{eq:l_acc}
    \frac{L_{\rm{acc}}}{\lambda} = 0.78 \frac{a_0^{2/3}}{\varepsilon_{cr}^{5/3} }  \left(\frac{n_e}{n_c}\right)^{2/3}.
\end{equation}

As electrons are accelerated, their energy increases and the laser pulse is progresivelly absorbed. Laser pulse absorption occurs both through energy transfer to accelerated electrons and through gradual erosion of the pulse front due to local pump depletion \cite{decker1996}. As the laser excites a nonlinear wake during propagation, it creates a sharp density spike at the front of the laser pulse. This spike causes localized laser downshift, resulting in the leading edge etching backwards. The etching velocity can be estimated as $v_{\rm etch} / c = \omega_p^2 / \omega_0^2$, which means that the pulse front velocity is the linear group velocity minus the etching velocity \cite{decker1996}. Since we operate in a high laser intensity regime where $a_0 \gg 1$ combined with low plasma density, the group velocity can be approximated as $v_g \approx c$ due to the relativistic correction to the refractive index, considering $v_{ph} = \left(1 - (\sqrt{2}n_e)/(a_0n_c) \right)^{-1/2}$.
This means that the front velocity in our case is $v_{\rm{front}}/c \approx 1 - \omega_p^2 / \omega_0^2$. Furthermore, laser absorption at the front can result in steepening of the laser pulse \cite{vieira2010}.

After the acceleration stage, an overdense thin foil is placed in the laser propagation direction to reflect the laser and ensure a head-on collision of accelerated electrons with the remaining unabsorbed laser pulse. The positioning of the foil is crucial in this scenario. If the foil is placed too close to the edge of the gas target, the acceleration distance is too short, which is insufficient for accelerating electrons to high energies. If the foil is placed too far inside, the laser pulse might become significantly absorbed, leaving only a small fraction of the initial laser energy for the collision. This consequently results in low pair production, despite electrons being accelerated to high energies.

The description of electron motion in an electromagnetic field, when radiated energy influences the trajectory, can be classified based on the dimensionless invariant $\chi$ parameter. It determines whether interactions are classical or in the QED regime, and for electrons it is defined as

\begin{equation}\label{eq:chi}
\chi_e = \frac{1}{E_s} \sqrt{\left( \gamma \vec{E} + \frac{\vec{p}}{m_ec}\times \vec{B} \right)^2 - \left( \frac{\vec{p}}{m_ec} . \vec{E} \right)^2} ,
\end{equation}

where $E_s$ is called Schwinger field \cite{schwinger1951}. It corresponds to the field that performs a work equal to electron rest energy $m_ec^2$ over a Compton length $\lambda_C = \hbar /m_ec$ and it corresponds to the intensity $\sim 10^{29}~\rm{W/cm^2}$. It can be defined as

\begin{equation}\label{eq:sch_field}
E_s = \frac{m_e^2c^3}{e \hbar}.
\end{equation} 

Generally speaking, electron trajectories can be treated within the framework of classical electrodynamics for $\chi<10^{-2}$, where the most used approach is Landau-Lifshitz (LL) radiation reaction that is treated as an additional damping force to the Lorentz force

\begin{equation}\label{eq:lorentz_force}
\frac{d\vec{p}}{dt} = e \left( \vec{E} + \frac{\vec{p}}{\gamma m_ec} \times \vec{B} \right) + \vec{F_{RR}}.
\end{equation}

For $\gamma \gg 1$, the standard LL formula can be approximated as \cite{landau1975}

\begin{equation}\label{eq:ll}
\vec{F_{RR}} = - \frac{2}{3} \frac{e^2\alpha \gamma \vec{p}}{m_e^3c^4}\left[ \left( \vec{E} + \frac{\vec{p}}{\gamma m_e} \times \vec{B}\right) ^2 - \left(\frac{\vec{p}}{\gamma m_e c} . \vec{E} \right)^2\right],
\end{equation}

where $\alpha$ is a fine structure constant. In the proposed scheme, the acceleration of electrons is modeled using LL formula, because accelerated electrons are co-propagating with the laser and $\chi<10^{-2}$. However, after the reflection, the value of $\chi$ rapidly increases and the quantum description is needed.

For $\chi > 10^{-2}$, the quantum nature of the interaction starts to become significant and stochastic effects appear. Instead of electrons continuously losing energy by emitting many photons with $\hbar \omega \ll m_e c^2$, high-energy photons are emitted via nonlinear Compton scattering with the differential probability $d^2P/dt, d\chi_\gamma$. After a photon is emitted via nonlinear Compton scattering, it can decay into an electron–positron pair with the differential probability $d^2P/dt, d\chi_e$ \cite{nikishov1967}. For compactness, the rates used in simulations are listed in Appendix \ref{ap:rates}.

The number of generated positrons can be estimated using an analytical expression derived by Blackburn et al. \cite{blackburn2017}, where the number of generated positrons is obtained for the scenario of a relativistic electron interacting with a laser pulse of given intensity and Gaussian envelope in the longitudinal direction. It provides the number of generated pairs by a single electron as a function of electron energy, laser pulse intensity, and pulse duration, assuming a plane wave in the transverse direction. The description can be extended to a realistic Gaussian beam geometry \cite{amaro2021,amaro2024}.

Number of pairs produced per electron can be expressed as

\begin{equation}\label{eq:pair_per_ele}
N_{+} \simeq 
\frac{3\sqrt{\pi} P_{\pm}(\omega_c) \chi_{c,\mathrm{rr}}}{\sqrt{2}}
\frac{(\gamma_0 m - \omega_c)^2}{\gamma_0 m}
\left. \frac{dN_{\gamma}}{d\omega} \right|_{\omega = \omega_c},
\end{equation}

where the term $P_{\pm}(\omega_c)$ represents the probability of emitting a photon of frequency $\omega_c$, $\chi_{c,\mathrm{rr}}$ is the recoil-corrected quantum parameter and the term $dN_\gamma / d\omega$ is the value of emitted photon distribution at $\omega=\omega_c$. For more details see \cite{blackburn2017}.

As a consequence of continuous electron loading during the DLA acceleration, energy spectrum is broad and the single-energy scaling needs to be generalized. Therefore, the number of pairs generated during the interaction of an electron population with distribution function $dN/dE$ can be estimated by the integral

\begin{equation}
N_{+}^{total} = \int_{E_{min}}^{E_{max}} N_+ \frac{dN}{dE}dE.
\end{equation}

All the scalings presented so far work well under the simplified assumptions used for their derivation. In reality, numerous nonlinear effects associated with laser propagation, direct laser acceleration, and pair creation lead to discrepancies between analytical models and realistic scenarios.

First of all, the propagation of a laser pulse is associated with relativistic self-focusing and laser diffraction. This leads to variations in laser intensity during propagation, making it impossible to maintain a constant spot size. Furthermore, curved laser wavefronts and a Gaussian transverse envelope cause electron acceleration to be less efficient compared to the analytical model, with longer acceleration distances needed than predicted by Eq. (\ref{eq:l_acc}). As the laser pulse shortens due to front depletion, electrons accelerated at the pulse front escape during the acceleration process, which lowers the beam charge and the maximum electron energy. As these electrons move ahead of the ion channel created by the laser, they no longer experience attractive transverse forces and escape the interaction region.

In this work, we place the reflecting foil at a distance $L_{\rm{acc}}$ from the onset of acceleration. Even though this distance should be sufficient for electrons to reach the energy $\gamma_{\rm{max}}$, due to the above-mentioned realistic features of laser propagation in the self-guided regime, electron energies are expected to be lower. However, it has previously been shown that cut-off energies can be in excellent agreement with theoretical predictions if the laser is externally guided \cite{babjak2025}.

Estimating the number of generated pairs requires knowledge of the electron energy distribution, as well as the laser duration and intensity at the moment of scattering. The shape of the distribution function in the regime where electrons are accelerated by DLA in low-density gas is challenging to predict and will be estimated based on simulation results. The laser duration can be estimated as $L(t)\approx L_0 - v_{\rm{etch}} t$ at early stages of the interaction. However, interaction of the laser front with the accelerated electron bunch effectively slows down the depletion, as discussed below. Furthermore, laser intensity evolves during propagation, making it impossible to predict the exact value of the peak laser intensity at the moment of reflection. Nevertheless, all necessary laser and electron properties can be extracted from simulation results.

The scaling for the number of pairs also assumes a Gaussian longitudinal shape of the laser pulse, with electrons entering the pulse from the outside. However, in our case electrons are located inside the laser pulse during acceleration, the laser front is steepened without a Gaussian entrance, and electrons are distributed throughout the pulse. As a result, the interaction time with the laser differs depending on their position inside the pulse. It is therefore necessary to be aware of the differences between the assumptions of the analytical scaling and the realistic interaction scenario. Even though it is not perfectly suited to provide exact predictions for the number of generated pairs in our scenario, it is useful for qualitative estimates and, as we will show, performs well despite the above-mentioned limitations.

\section{Laser pulse depletion during the direct laser acceleration of electrons}\label{s:depletion}

\begin{figure}
    \centering
    \includegraphics[width=0.99\linewidth]{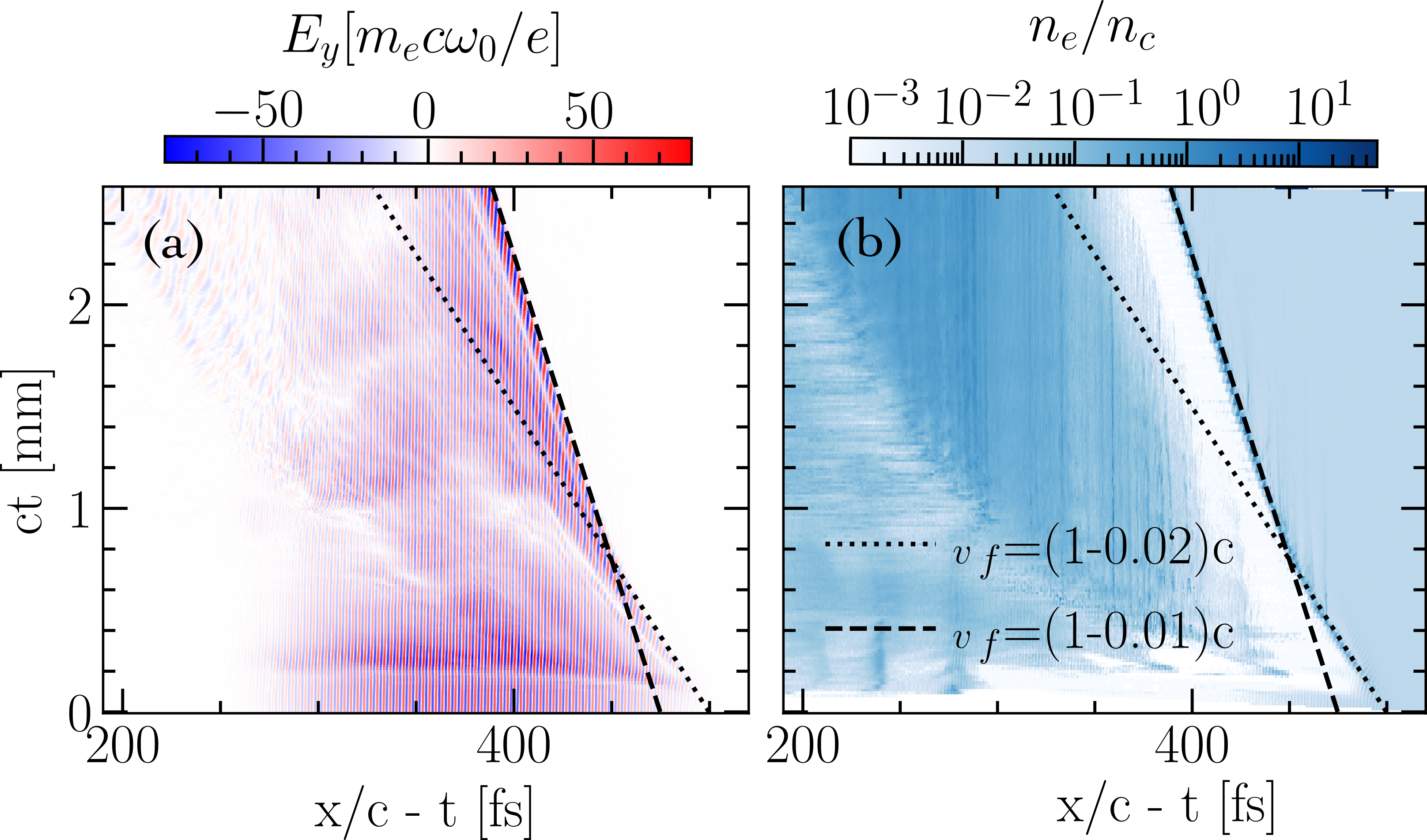}
    \caption{Temporal evolution of the: (a) Transverse component of the laser field (b) Electron plasma density. For each timestep shown, a lineout is presented considering a simulation window moving at the speed of light. In the beginning, the front etching velocity matches the theoretical prediction, slowing down after the propagation for $\approx$ 800 microns. The data shown is taken from quasi-3D PIC simulations, with a 6 PW laser propagating through the plasma with the density of $n_e/n_c = 0.02 $.  }
    \label{fig:lineouts}
\end{figure}

The evolution of the laser shortening process due to local pump depletion is illustrated in Fig. \ref{fig:lineouts}, which shows a simulation of a 6 PW laser pulse propagating through plasma with constant density $n_e = 0.02 n_c$. At the beginning of the propagation, the laser pulse front propagates with a velocity of $(1-0.02)c$, which is in agreement with the analytical prediction. The lineouts are taken from a moving-window simulation that propagates at the speed of light $c$. The laser edge is moving backwards, effectively shortening the laser pulse.

After 0.8 mm of propagation, the laser front velocity changes, slowing down the laser depletion. This can be caused by the interaction with the accelerated electron beam loaded inside the laser pulse, potentially influencing the structure of the density spike. However, more detailed analysis and theoretical exploration of other effects that can affect the etching velocity are needed to fully confirm this.

Additionally, lower-frequency waves are clearly visible in the top left corner of Fig. \ref{fig:lineouts}(a), which originate from laser frequency downshift at the front. At the beginning of the interaction ($ct < 0.2$ mm), electrons inside the laser pulse are fully evacuated due to the ponderomotive force. However, electrons are clearly injected slightly later, still at early stages of the interaction.

\begin{figure}
    \centering
    \includegraphics[width=0.5\linewidth]{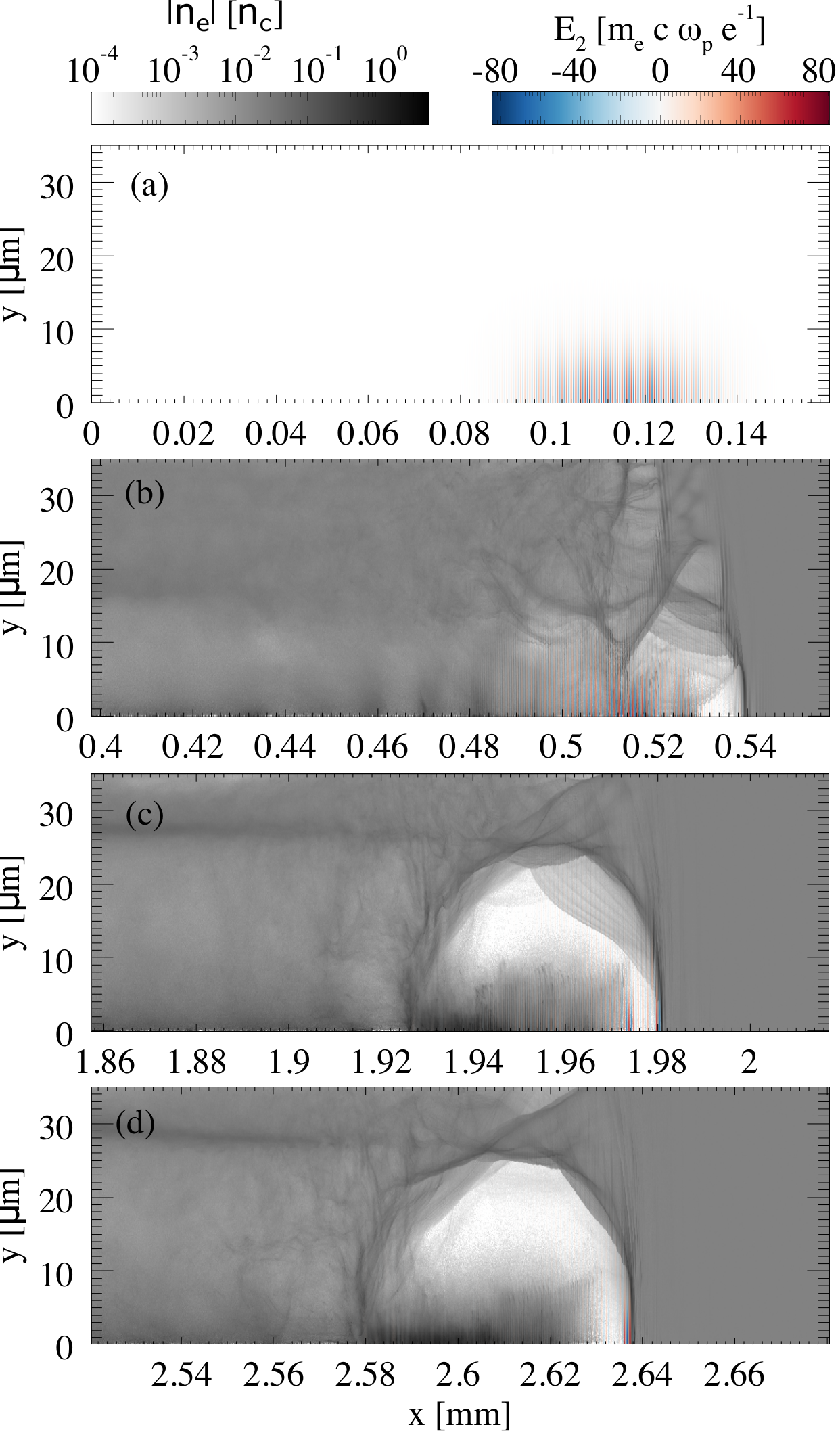}
    \caption{ The dynamics of a 6 PW-laser propagating through the constant density plasma at 0.02 $n_c$ . In panel (a), the laser is initialized in vacuum. Panels (b-d) demonstrate the laser propagation in self-guided regime for 2.5 mm during which electrons get accelerated.  }
    \label{fig:2d_propagation}
\end{figure}

\begin{figure}
    \centering
    \includegraphics[width=0.5\linewidth]{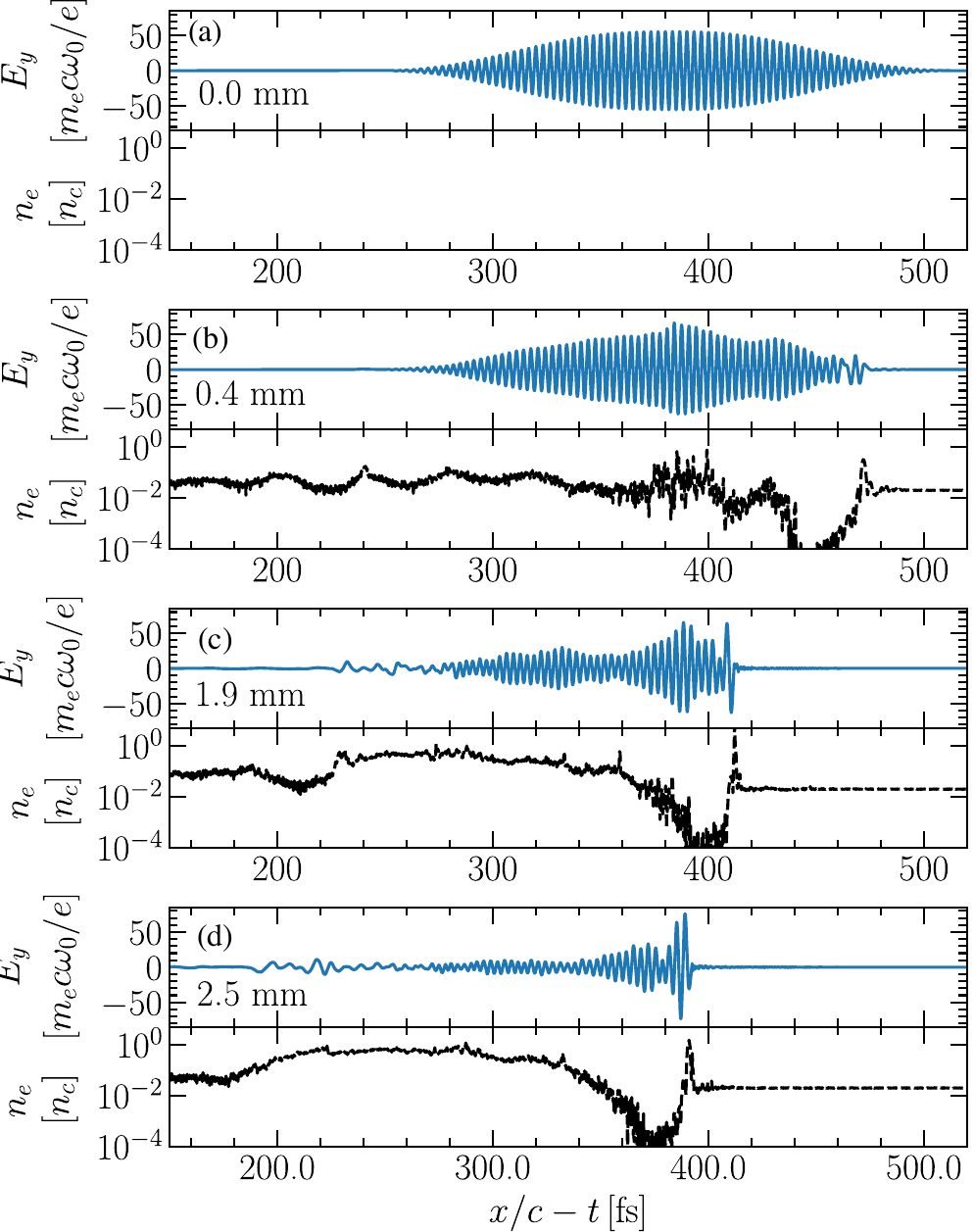}
    \caption{Lineouts of the laser transverse electric field and the electron plasma density after different distances of propagation inside the plasma. A 6 PW laser propagates through the plasma with the density $n_e/n_c = 0.02$. A steep density spike at the laser front causes photon deceleration, visible as a long-wavelength laser field at the back. Laser front steepening due to local pump depletion is also present. }
    \label{fig:evolution}
\end{figure}

A detailed look at the evolution of the on-axis electron density and laser pulse is shown in Fig. \ref{fig:2d_propagation} and the 2D equivalent in Fig. \ref{fig:evolution}. Panels in both figures correspond to the same times during the interaction. At the beginning of the propagation, at $ct=0$ mm, the unperturbed laser pulse initialized in vacuum is shown. After 1 mm of propagation, electron beam loading is clearly visible, with a sharp density spike at the front of the laser pulse. Furthermore, the intensity at the laser front is steepened as expected, with self-modulation also being present.

As the laser pulse propagates further inside the gas target, it becomes shorter, with the electron bunch also slightly shifting backwards. After 2.5 mm of propagation, the laser pulse is almost fully absorbed, with only a small fraction of the electron bunch overlapping with the laser.

\section{Properties of accelerated electrons and secondary positrons}\label{s:particle_properties}

The QED regime of interaction can be defined by the condition on the quantum nonlinearity invariant $\chi_e > 1$. To explore the regime of interest, we estimate the minimal laser requirements needed to reach this regimeby running a set of simulations for 2, 4, 6, 8, and 10 PW lasers with a duration of 150 fs. Simulations were performed in quasi-3D geometry using OSIRIS \cite{fonseca2002,davidson2015}. Simulation details are listed in Appendix \ref{ap:sim_params}.

A background hydrogen plasma was initialized with constant density along the propagation direction, with no preformed guiding channel used. The laser spot size and field amplitude $a_0$ were chosen according to the optimal focusing strategy presented in our previous work \cite{babjak2024}. However, without a preformed guiding structure, the laser pulse spot size changes as a result of relativistic self-focusing and oscillates during the interaction. This means that a fully optimal regime is not reachable because it is not possible to sustain a constant matched spot size. As a result, electron energies are expected to be lower than predicted by Eq. (\ref{eq:gmax_resonant}).

Properties of the initialized plasma and the laser are summarized in Table \ref{tab:comptons_sim_params}. An overdense thin foil with a density of 435 $n_c$ is placed at a distance $L_{\rm{foil}}$ inside the gas. The distance is chosen according to an estimate that determines the distance needed to achieve the maximum energy given by Eq. (\ref{eq:l_acc}). This is a reasonable choice for proof-of-principle simulations, where electrons should be accelerated to sufficiently high energies while the laser pulse is still not fully absorbed. Possible strategies for choosing the optimal position of the reflecting foil will be further discussed in Section \ref{s:mirror_position}.

\begin{table*}
\centering

\begin{tabular}{c|c|c|c|c|c}
  Laser power  & Laser energy & $W_0$  & $a_0$ &  $n_e$  & $L_{\rm{foil}}$ \\

    (PW) &  (J) & ($\mu m$) &  &  ($n_c$) &  (mm)\\
 \hline\hline
  2 &  234 &  8.0 & 38.2 &  0.05 & 1.1 \\
  \hline
  4 &  468 & 8.5 & 50.8 & 0.02 & 2.2 \\
  \hline
  6 &  702 & 9.0 & 58.7 &  0.02 & 2.55 \\
  \hline
  8 & 936 & 9.0 & 67.9 &  0.03 & 2.0 \\
  \hline
  10 &  1170 & 9.0 & 75.9 &  0.03 & 2.1 \\

\end{tabular}

\caption{ Laser and plasma parameters used in simulations chosen based on the optimal focusing strategy. $W_0$ is laser waist at focus, $a_0$ is the peak laser normalized amplitude, $n_e$ is the plasma density and $L_{\rm{foil}}$ is the propagation distance after which the overdense foil was placed to ensure the laser reflection.  }\label{tab:comptons_sim_params}
\end{table*}

\begin{figure}
    \centering
    \includegraphics[width=0.6\linewidth]{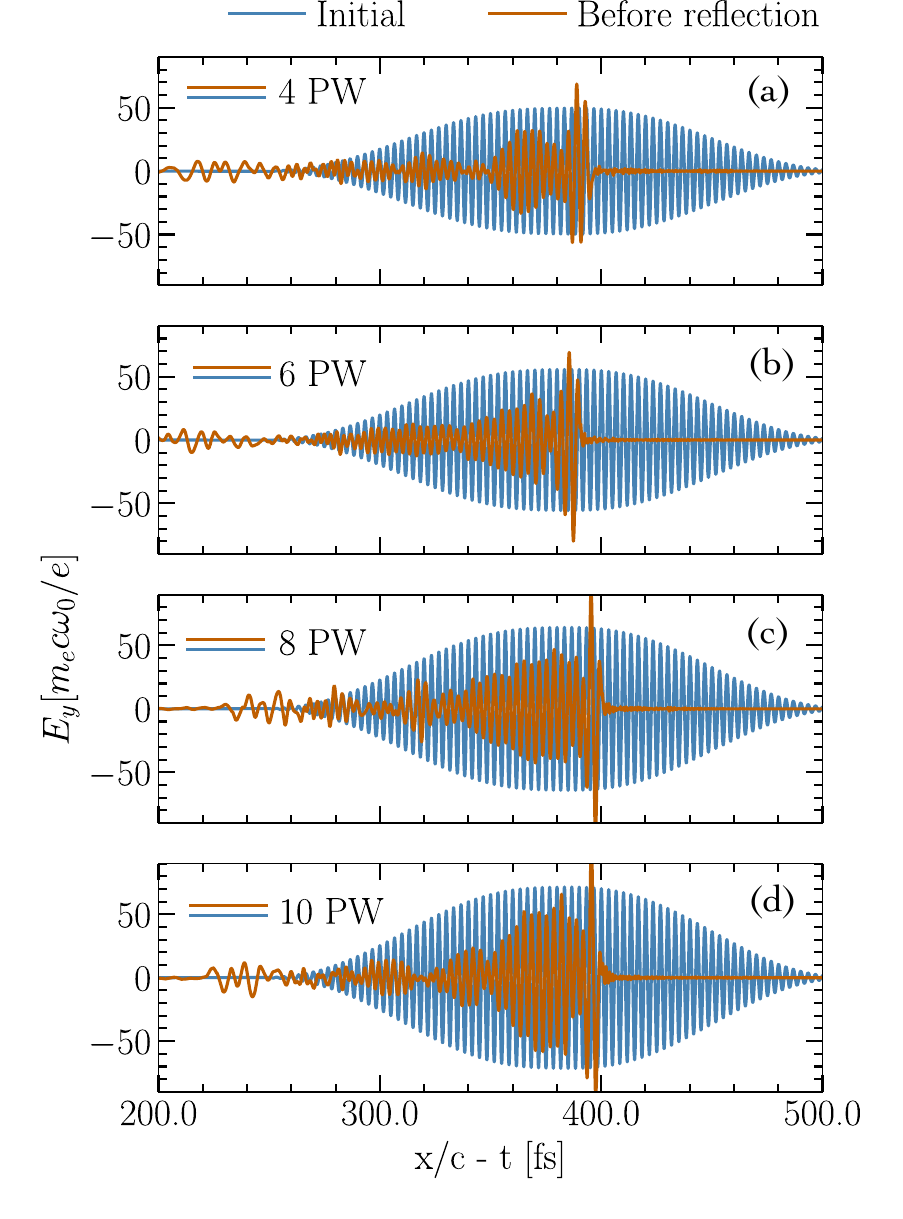}
    \caption{Comparison of initial laser pulse shape with the laser pulse shape immediately before the reflection from the overdense target for different laser powers.  }
    \label{fig:laser_power_scan}
\end{figure}

The comparison of the laser pulse shape between its initialization (before interaction with the gas) and just before reflection from the thin foil is shown in Fig. \ref{fig:laser_power_scan}. The shape of the envelope is significantly altered by the nonlinear interaction with the gas. In all cases, the laser is shortened with a steep front, sustaining a maximum laser field amplitude comparable to the initial one. A short high-intensity spike is present in all cases. Due to its duration, it has a negligible impact on the number of generated pairs; however, it affects the maximum value of $\chi$ for electrons interacting with the reflected laser.

Spectra of accelerated electrons after the propagation distance $L_{\rm{foil}}$ are shown in Fig. \ref{fig:spectra}(a). The spectra have a shape typical for efficient DLA, consisting of a waterbag-like part followed by a Boltzmann-like high-energy region. Cut-off energies are lower compared to the values predicted by Eq. (\ref{eq:gmax_resonant}) for several reasons. First, electron energies could be higher if propagation were sustained over a longer distance. However, this would result in more laser energy being absorbed, which would not be beneficial for pair creation due to the laser pulse becoming too short. Furthermore, oscillations of the laser pulse spot size cause the acceleration to be effectively slower compared to the plane-wave-based description used for the derivation of the acceleration distance needed to reach the maximum energy. The total accelerated charge of electrons with energies above 500 MeV is above 100 nC for all the cases, namely 100 nC for the 4 PW laser, 140 nC by the 6 PW alser, 113 nC by 8 PW and 165 nC for the 10 PW case.

After the collision with the reflected pulse, spectra of generated positrons are shown in Fig. \ref{fig:spectra}(b). The number of positrons increases with laser power as expected, with 0.03 nC of positrons created by the 4 PW laser and 2.1 nC by the 10 PW laser. It is worth noting the promising nonlinear scaling with increasing laser power, resulting in the creation of two orders of magnitude more pairs for 2.5 times higher power. The reason is the strong nonlinearity of the Breit–Wheeler pair creation rate with respect to $\chi$.

\section{Laser requirements for reaching QED regime for prolific pair creation}\label{s:qed}

The increasing number of generated pairs scaling with laser power occurs due to the increasing number of electrons interacting with the laser pulse that reach $\chi$ values greater than 1. For a head-on collision between a linearly polarized laser pulse and a relativistic electron, $\chi_e$ can be estimated as $\chi_e \approx 3.4 \times 10^{-6} a_0 \gamma_e$. 

In the optimized DLA regime, where the spot size is chosen to match the resonant oscillation amplitude of electrons, it is possible to estimate the cut-off energy. First, the combination of laser spot size and $a_0$ is determined by Eq. (4) in Ref. \cite{babjak2024} for an assumed background plasma density $n_e$ and laser power $P$. The values of $a_0$ and $n_e$ are then substituted into the prediction for the highest achievable energy in the DLA regime given by Eq. (\ref{eq:gmax_resonant}).

Note that the maximum electron energy achievable for a given laser power does not depend strongly on plasma density in the range $10^{18}~\rm{cm^{-3}}$–$10^{20}~\rm{cm^{-3}}$. However, the DLA optimum is achieved with higher $a_0$ (smaller $W_0$) at higher densities, which should be beneficial for pair creation. Nevertheless, we omit studying the pair generation scheme in the density range around $\approx 0.1n_c$ or higher. First, laser propagation is less stable, which may complicate the analysis. Furthermore, higher densities combined with relativistic laser intensities can trigger acceleration in the radiation reaction–dominated regime, making the interaction even more complex. However, using higher plasma densities can be a pathway toward reaching the QED regime with even lower laser power and will be the subject of future work.

To estimate the maximum $a_0$ at the moment of reflection, we use the initial values at focus for simplicity. This is a very rough estimate, but determining the real value analytically is challenging due to the nonlinearity of the interaction. First, $a_0$ varies during acceleration due to self-focusing and defocusing in plasma. Furthermore, in all cases the highest field amplitude is present in the high-intensity spike at the front of the laser pulse (see Fig. \ref{fig:laser_power_scan}), reaching values up to twice as high as the rest of the laser.

\begin{figure}
    \centering
    \includegraphics[width=0.99\linewidth]{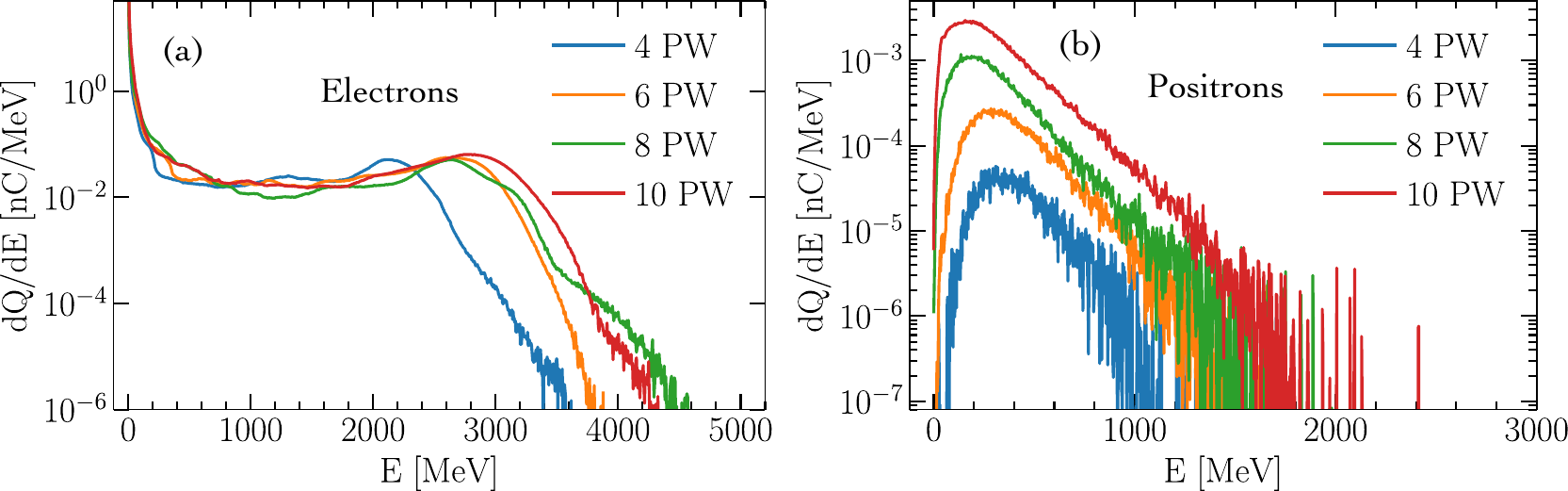}
    \caption{(a) Electron energy spectra before interaction with the overdense foil (b) Generated positrons energy spectra right after leaving the overdense target. }
    \label{fig:spectra}
\end{figure}

\begin{figure}
    \centering
    \includegraphics[width=0.99\linewidth]{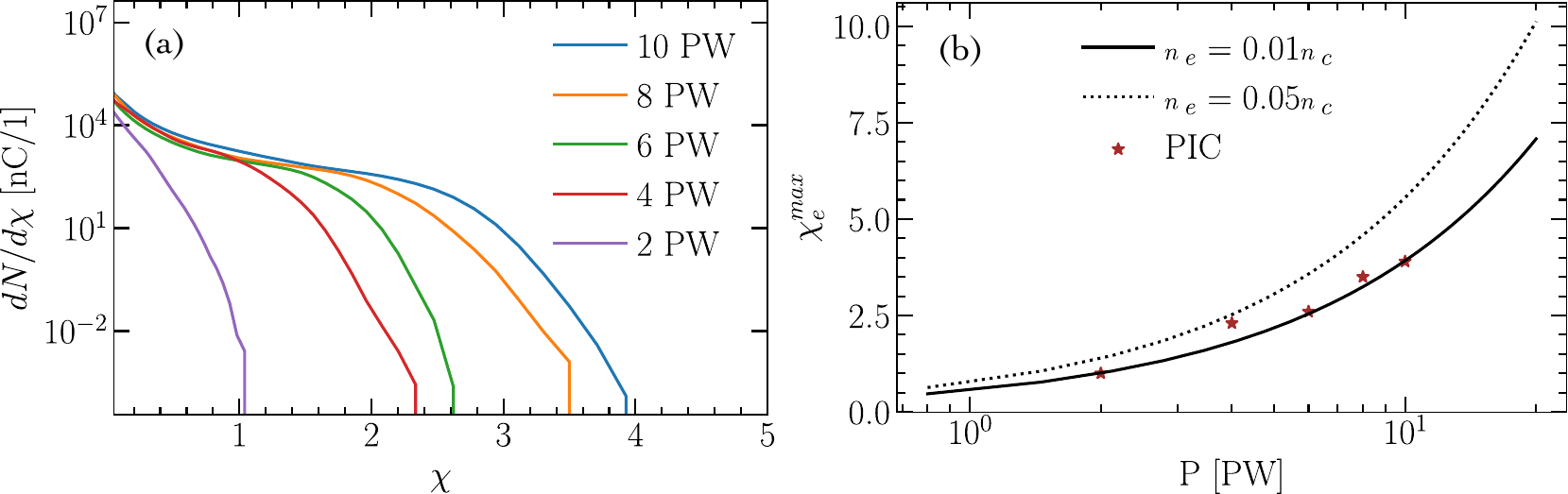}
    \caption{The electron quantum parameter $\chi_e$ at the time of photon emission. (a) Cumulative $\chi_e$ distribution over the entire simulation for lasers of different powers. (b) The maximum $\chi_e$ as a function of laser power: the analytical estimate compared with the value obtained from PIC simulations. Here, Eq. (\ref{eq:gmax_resonant}) is used to predict the electron energies after acceleration and Eq. (4) in Ref. \cite{babjak2024} was used to choose the optimal laser-plasma parameters for acceleration maximizing $\gamma_e$ and consequently $\chi_e$. }
    \label{fig:chi}
\end{figure}

Using this strategy to estimate the highest $\chi_e$ achievable with, for example, a $10~\mathrm{PW}$ laser, we take $a_0 \approx 76$ and $\gamma_{\rm max} \approx 17\,600$, which yields an estimate of $\chi_e^{\rm max} \approx 4.5$. This value is in good agreement with that observed in the PIC simulations, see Fig. \ref{fig:chi}(a). 

Nevertheless, it is important to emphasize that this estimate is only approximate and relies on simplifying assumptions that must be treated with caution. In particular, discrepancies can arise both from an overestimation of the maximum electron energy and from variations of the effective laser amplitude $a_0$ during propagation. For the $10~\mathrm{PW}$ case, the electron energy observed in the simulations is lower than the analytically predicted maximum, with $\gamma \simeq 9,000$ (see Fig.~\ref{fig:spectra}(a)), while the effective $a_0$ experienced by the electrons is significantly enhanced compared to its initial focal value, reaching peak values of about $115$.

Since $\chi_e$ depends on both the electron Lorentz factor and the local field strength, these two effects partially compensate each other. As a result, despite the strongly nonlinear nature of the interaction, the estimate provides a reasonable prediction of the maximum $\chi_e$ achievable in this regime.

The cumulative distribution of $\chi_e$ for electrons during photon emission by nonlinear Compton scattering in the field of the reflected laser is shown in Fig. \ref{fig:chi}(a). With increasing laser power, more electrons radiate with $\chi_e > 1$ due to higher electron energies and higher laser amplitude $a_0$. The 2 PW simulation demonstrates that this is the threshold laser power needed to reach the limit of $\chi_e = 1$. Even though pair creation is possible for $\chi_e < 1$, we would need to run simulations with a higher number of macroparticles to resolve the spectrum of generated pairs with sufficient accuracy. Therefore, the analysis of pair creation by the 2 PW laser is omitted due to low pair statistics in this simulation.

The maximum $\chi_e$ achieved in each case is extracted and plotted in Fig. \ref{fig:chi}(b). It is compared with the estimate for the maximum $\chi_e$ calculated for $n_e = 0.01 n_c$ and $n_e = 0.05 n_c$. As discussed above, $\chi_e$ can potentially be increased by using higher plasma density, but for simplicity we restrict ourselves to densities on the order of $0.01 n_c$. Higher densities are accompanied by more unstable laser pulse propagation, faster laser depletion, and strong energy loss due to radiation reaction \cite{jirka2020}. We find good agreement with the prediction, despite uncertainties regarding the estimates of $\gamma_e$ and $a_0$.

A summary characterizing the interaction for different laser powers is given in Table \ref{tab:pairs_summary}. Note the favorable scaling of the number of generated pairs per initial laser energy. The reason is that the number of generated pairs increases nonlinearly with laser intensity and electron energy. The reason the number of emissions at $\chi > 1$ exceeds the number of accelerated electrons is that a single electron can emit several times during its interaction with the reflected laser before losing significant energy.

\begin{table*}
\centering

\begin{tabular}{c|c|c|c|c}
  Laser power & Laser energy  & Emissions $\chi_e>1$ & Generated pairs  & Positrons / Laser energy  \\

    (PW) &  (J) & (nC) &   (nC) &  (pC/J)\\
 \hline\hline
  4 &  468 & 190 & 0.03 & 0.06 \\
  \hline
  6 &  702 & 420 & 0.19 &  0.27 \\
  \hline
  8 & 936 & 680 & 0.73 &  0.78 \\
  \hline
  10 &  1170 & 1010 & 2.12 &  1.81 \\

\end{tabular}

\caption{ Properties characterizing the nonlinear Compton scattering pair creation for different laser parameters in optimal DLA conditions. }\label{tab:pairs_summary}
\end{table*}

\section{Analytical estimate for the number of generated pairs}\label{s:number_estimate}

For the setup based on the combination of DLA acceleration and reflection from the plasma mirror, several assumptions need to be made. First of all, the shape of an electron spectrum cannot be exactly reproduced for the purpose of the analytical estimate. Based on results shown in Fig. \ref{fig:spectra} (a), we approximate it as a constant function with a value of 0.025 nC/MeV up to the maximum energy $f \times \gamma_{\rm{max}}$. The factor $f$ is used according to observations of the electron cut-off based on simulation results (see Fig. \ref{fig:spectra}). We estimate the cut-off coefficient to be in the range $f=[0.4,0.5]$. The value $E_{\mathrm{min}}$ of the assumed distribution is not important in this case because low-energy electrons do not contribute to pair creation. The value of $\gamma_{\rm{max}}$ is calculated using the same procedure as we used for the estimate of $\chi_e^{\rm max}$, assuming background plasma density $n_e = 0.03 n_c$. The difference between predicted maximum energy at lower or higher densities in optimal conditions and chosen energy range is negligible, much smaller than the uncertainties in the assumed electron spectrum shape.

We can use Fig. \ref{fig:laser_power_scan} to estimate the laser pulse duration, which is also needed to predict the number of pairs. The initial duration of 150 fs was decreased due to local pump depletion to approximately 50 fs, which we use for the scaling, along with the initial value of $a_0$. Even though the spot size of the laser oscillates during propagation, the average value or amplitude of these oscillations cannot be predicted due to the complexity of the interaction between the laser and the plasma.

\begin{figure}
    \centering
    \includegraphics[width=0.7\linewidth]{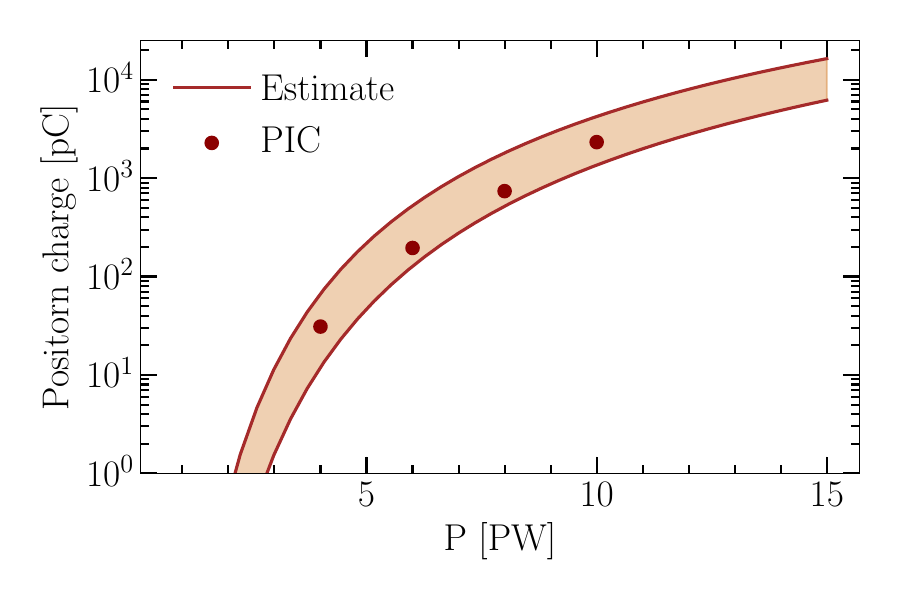}
    \caption{The number of pairs generated during the single-laser DLA-based scheme. The measured values in PIC simulation are compared with the theoretical estimates. To account for nonlinear effects in the DLA stage, we defined the window of expected values assuming a range of achieved electron energies between 40\% and 50\% of the allowed maximum $\gamma_{\rm{max}}$  for the ideal case.  Laser duration at reflection is 50 fs and background plasma density used to determine optimal acceleration parameters is 0.03 $n_c$. }
    \label{fig:pairs_number}
\end{figure}

The comparison of the semi-analytical estimate with the number of pairs obtained in PIC simulations is shown in Fig. \ref{fig:pairs_number}. Despite numerous assumptions, the number of pairs observed in simulations falls well within the range of uncertainty. Note the sharp decrease in the number of generated pairs for laser powers <2 PW, due to the difficulty of reaching the regime $\chi_e>1$. According to the scaling, it should be possible to reach approximately 10 nC of positrons with a 15 PW laser.

\section{Optimal positioning of the reflecting mirror}\label{s:mirror_position}

There are several possibilities to optimize the process and increase the number of produced positrons. By introducing an external neutral guiding channel (e.g., using a prepulse or a discharge), the energy of electrons can be significantly increased due to stable laser propagation and sustaining the optimal spot size during acceleration, which is difficult to achieve in the self-guided regime for these parameters.

Another important variable in the proposed setup is the position of the reflection foil. If it is positioned too early during the interaction, electrons cannot become energetic enough because of the short acceleration distance. On the other hand, if the reflection occurs too late, the laser can be significantly depleted, resulting in too short laser duration, insufficient for pair creation. The highest number of pairs is expected if the largest possible number of ultra-relativistic electrons interacts with the longest reflected pulse.

To test how much the number of generated pairs can be influenced by foil positioning and to understand the optimal position, we ran a simulation of an externally guided 10 PW laser pulse. The laser amplitude is defined as $a_0(x)=a_0 \sin^2[\pi x / (2T_L)]$ in the longitudinal direction, with $T_L=150~\rm{fs}$, $W_0=10~\mu$m, and $a_0=68$. The FWHM of the laser intensity is approximately 110 fs, with a total energy of 1125 J. A parabolic guiding channel with a radius of 20 $\mu$m was initialized with the density on the channel axis $n_e=0.01~n_c$ and $n_e=0.1~n_c$ on the channel wall.

\begin{figure}
    \centering
    \includegraphics[width=0.99\linewidth]{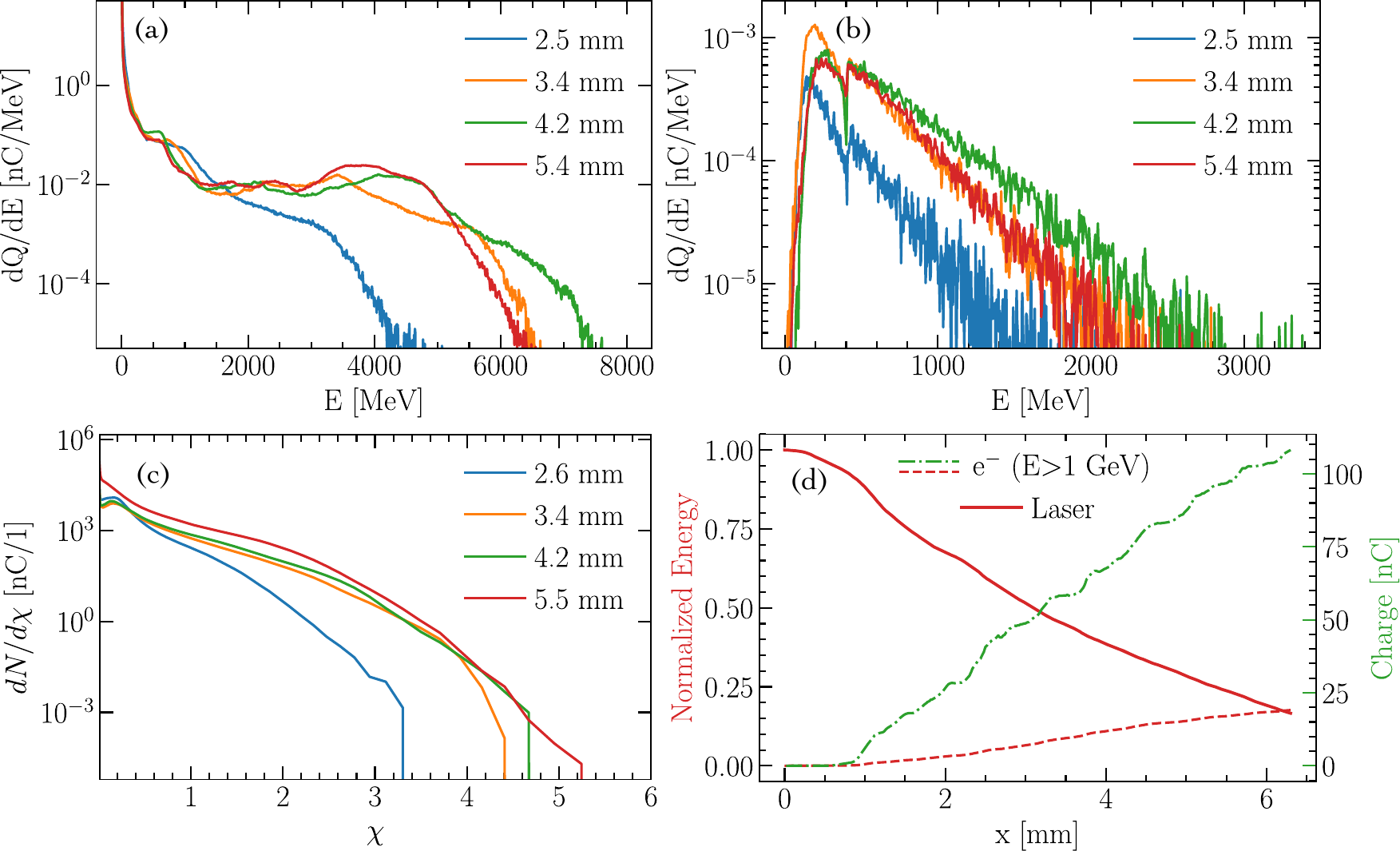}
    \caption{The effect of the overdense foil positioning on pair creation for an externally guided 10 PW laser. (a) Energy spectra of electrons before the reflection from the foil. (b) Positron spectra for various positions of the foil inside the gas target. (c) Cumulative $\chi_e$ distribution of electrons emitting hard photons during the collision of the reflected laser pulse. (d) Evolution of the laser energy, energy of electrons above 1 GeV and total accelerated charge as a function of the propagation distance. The energy axis is normalized to the initial laser energy of 1125 J. }
    \label{fig:scan_position}
\end{figure}

The general picture of the interaction is shown in Fig. \ref{fig:scan_position} (d). Laser energy is absorbed, with only 20 percent of the energy remaining after 6 mm of propagation. The energy transferred from the laser to electrons above 1 GeV ($\chi_e \simeq 0.5$) increases linearly with the propagation distance due to continuous injection and acceleration, resulting in a 20 percent conversion efficiency at the end of the interaction. The total charge of electrons above 1 GeV also increases linearly with propagation, reaching 100 nC.

The foil used to reflect the laser is positioned after four different propagation distances: 2.5 mm, 3.4 mm, 4.2 mm, and 5.4 mm. Electron spectra before the laser reflection are shown in Fig. \ref{fig:scan_position} (a). The energy cut-off can differ significantly: if the foil is placed too soon, the energy cut-off is only 4 GeV, whereas 7 GeV electrons are observed after 4.2 mm of propagation. Positron spectra are shown in Fig. \ref{fig:scan_position} (b). Discontinuities in the spectrum at around 400 MeV are numerical artifacts and do not influence the creation rate or the number of generated positrons, as discussed in more detail in Appendix \ref{ap:numerics}. The highest-energy positrons are created in the scenario when electron energies are the highest. The cumulative $\chi_e$ distribution in Fig. \ref{fig:scan_position} (c) demonstrates that, in all cases, electrons interacted with the laser in a similar regime, with the maximum $\chi_e$ reaching 5.

By increasing the distance at which the foil was placed, different numbers of positrons were created. For the distances 2.5 mm, 3.4 mm, 4.2 mm, and 5.4 mm, the number of pairs was 0.26 nC, 0.91 nC, 0.89 nC, and 0.73 nC, respectively. This means that, in the best-case scenario, the efficiency of pair creation was 1.81 pC per Joule of laser energy. This is lower compared with the 10 PW scenario in the self-guided regime, which may seem counter-intuitive. However, this is caused by electrons being accelerated to higher energies ($\approx$7 GeV instead of $\approx$4 GeV) in the externally guided scenario, which generated more energetic positrons. While in the optimized case positron spectra extended up to almost 3 GeV, in the self-guided simulation electron energies reached only up to 2 GeV. Furthermore, a slight advantage in the self-guided regime could be due to the use of a higher density, resulting in higher laser $a_0$ at reflection. Understanding how much the density increase can benefit pair creation will be the topic of future work. It is currently beyond the scope of this study, as it requires a detailed understanding of DLA beyond the assumption of the laser phase velocity $v_{\rm ph}=c$, as well as laser pulse propagation and depletion in this high-density regime, which is generally associated with more unstable propagation.

To predict the optimal position for the overdense foil, several complementary questions need to be answered. First, a more accurate model for etching velocity needs to be developed to capture the later stages of laser propagation. Furthermore, the shape of the electron distribution function and its evolution in time need to be better understood in conditions without a preformed guiding channel. Electron properties also strongly depend on whether acceleration occurs in a purely DLA regime or if additional acceleration by nonlinear LWFA is present. This depends strongly on the laser duration and needs further study. Only after laser absorption and acceleration are understood in greater detail, it is possible to analytically develop strategies to optimize the number of generated positrons. However, we have demonstrated that the scheme is robust and produces a high number of secondary particles over a wide range of foil positions, making it suitable for experimental verification.

\section{Conclusion}

In this work, we investigate a single-laser positron generation scheme based on the non-linear Breit–Wheeler process. After a laser pulse accelerates electrons, it is reflected from an overdense thin foil and interacts with the accelerated electrons in a head-on collision, resulting in prolific pair creation. We demonstrate that already a 2 PW laser pulse is sufficient to reach the quantum limit, where the quantum nonlinearity parameter $\chi_e>1$, already in a non-guided regime. A semi-analytical model is used to predict the number of generated positrons, agreeing with the rate of generating 1.8 pC/J of laser energy observed in self-consistent PIC simulations.

In a simulation of an externally guided 10 PW laser pulse, we see that it is possible to reach $\chi_e \approx 5$, with scalings predicting $\chi_e \approx 10$ for 20 PW laser pulses. This is not easy to achieve with a two-step colliding scheme, because electrons would lose energy before interacting with the center of the pulse. Here, this is a consequence of self-steepening in a plasma. The electrons first interact with the maximum laser intensity reflected from the foil. This can be guaranteed by a careful choice of setup parameters. The presented scheme provides an experimentally feasible platform to study strong-field QED phenomena on already existing laser facilities.

\begin{acknowledgments}
This work was supported by FCT Grants PTDC/FIS-PLA/3800/2021 \newline DOI: https://doi.org/10.54499/PTDC/FIS-PLA/3800/2021, FCT UI/BD/151560/2021 \newline DOI:https://doi.org/10.54499/UI/BD/151560/2021 and 2023.16184.ICDT \newline DOI: https://doi.org/10.54499/2023.16184.ICDT. We acknowledge use of the Marenostrum 5 (Spain) and Karolina (Czechia) supercomputers through EuroHPC awards. This work was supported by the Ministry of Education, Youth and Sports of the Czech Republic through the e-INFRA CZ (ID:90254). The authors acknowledge fruitful discussions with Óscar Amaro. 
\end{acknowledgments}

\appendix

\section{Particle-in-cell simulation parameters}\label{ap:sim_params}

In simulations where laser pulse interacted with the constant density plasma, following parameters were chosen. Laser pulse transverse profile is defined as $a_0(r) = a_0 \exp(-r^2/W_0^2)$, so the laser waist $W_0$ corresponds to the distance where the field amplitude is $1/e$ of the maximum. The longitudinal laser envelope is chosen as polynomial function $a_0(\tau)=a_0\times [10\tau^3-15\tau^4+6\tau^5]$ expressing the first half of the envelope where $a_0$ increases, where $\tau = t/T_L$ and $T_L$ is a laser duration in FWHM of the laser field amplitude. The second part is symmetric around the point of the highest intensity, but instead with a decreasing field amplitude. This means that if we refer to the laser pulse with a duration $T_L$, the interval [$t_0-T_L/2$;$t_0+T_L/2$] around the laser center at $t_0$ contains 94\% of the laser energy. For such chosen laser pulse, the total energy can be expressed as $E[\rm{J}]=0.78\times P[\rm{PW}]\times T_L[\rm{fs}]$. The grid was discretized using 60 cells per wavelength in the laser propagation direction and 21 cells per wavelength in the transverse direction. The first two azimuthal modes were used for simulations. They account for the axisymmetric self-generated channel fields (mode m = 0) and for the non-axisymmetric linearly polarized laser field (m = 1). 32 particles per cell were used both for ion and electron species. We assume laser wavelength $\lambda= 1~\mu$m and the timestep $\Delta t = 25.4~\rm{as}$. 

Externally guided 10 PW simulation was initialized in a following way. The grid discretization $N_x \times N_y=60 \times 20 (c/\omega_0)^2$ with the timestep of 25.48 as. Background plasma is initialized using $2\times 2 \times 8$ particles in $x,r$ and $\varphi$ direction respectively. Electromagnetic fields are pushed using dual solver that is numerically well suited for simulations of DLA acceleration in underdense plasmas. Particle pusher contains classical radiation reaction in the Landau-Lifshitz form. Simulation assumes laser wavelength $\lambda = 1 ~\mu$m. The 10 PW simulation contained a laser with normalized field amplitude $a_0 = 68 $ and spot size $W_0 = 10 ~\mu$m. Laser pulse transverse profile is defined as $a_0(r) = a_0 \exp(-r^2/W_0^2)$, so the laser waist $W_0$ corresponds to the distance where the field amplitude is $1/e$ of the maximum. In longitudinal direction the laser amplitude was defined as $a_0(x)=a_0 \sin^2[\pi x / (2T_L)]$ and the value $T_L=150 ~\rm{fs}$. The density in the simulation was initialized by a parabolic quasi-neutral guiding plasma channel with a transverse density profile in a form $(n_w-n_e)(r/R_{ch})^2 + n_e$, where $n_w$ is a channel wall density, $R_{ch}$ is a plasma channel radius and $n_e$ is a plasma density at the center of a channel, whilst $n_e = 0.01n_c$, $n_w = 0.1n_c$ and $R_{ch}=20~\mu$m.

\section{Differential probabilities of QED events}\label{ap:rates}

Instead of electrons continuously loosing energy by emitting many photons with $\hbar \omega \ll m_ec^2$, high energy photons are emitted via nonlinear Compton scattering with the differential probability \cite{nikishov1967}

\begin{equation}
    \frac{d^2 P}{dt\, d\chi_\gamma}
    = 
    \frac{\alpha m c^2}{\sqrt{3}\pi \hbar \gamma \chi_e}
    \left[
        \left( 1 - \zeta + \frac{1}{1 - \zeta} \right)
        K_{2/3}(\tilde{\chi})
        - 
        \int_{\tilde{\chi}}^{\infty} dx\, K_{1/3}(x)
    \right],
    \label{eq:diff_prob_rate}
\end{equation}
where $\tilde{\chi} = \dfrac{2\zeta}{3\chi_e(1 - \zeta)}$ and $\zeta = \dfrac{\chi_\gamma}{\chi_e}$.

The $\chi_\gamma$ is a quantum nonlinearity parameter for photons with the wave 4-vector $k^\mu$ instead of a particle 4-momentum $(E/c,\vec{p})$ defined as

\begin{equation}\label{eq:chi_photons}
\chi_\gamma = \frac{\sqrt{(\hbar k_\mu F^{\mu \nu})^2}}{E_sm_ec},
\end{equation}

where $F^{\mu \nu}$ is an electromagnetic tensor. The differential rate of pair production by a photon in a background electromagnetic field is given by \cite{nikishov1967} 
\begin{equation}
    \frac{d^2 P}{dt\, d\chi_e}
    =
    \frac{\alpha m^2 c^4}{\sqrt{3}\pi \hbar^2 \omega \chi_\gamma}
    \left[
        \left(
            \frac{\zeta^+}{\zeta^-} + \frac{\zeta^-}{\zeta^+}
        \right)
        K_{2/3}(\tilde{\chi})
        +
        \int_{\tilde{\chi}}^{\infty} dx\, K_{1/3}(x)
    \right],
    \tag{3.11}
\end{equation}
where $\tilde{\chi} = \dfrac{2}{3\chi_\gamma \zeta^+ \zeta^-}$ and $\zeta^+ = \dfrac{\chi_e}{\chi_\gamma} = 1 - \zeta^-$.

\section{Verification of the reliability of the numerical QED parameters on pair creation }\label{ap:numerics}

The QED module of OSIRIS contains a parameter called \textit{qed\_g\_cutoff}, which determines if particles is pushed by a pusher accounting for a quantum radiation reaction or by the classical Landau-Lifshitz (LL) pusher. For computational reasons, it is necessary to distinguish between the population of particles that should be treated by the QED or classical module. If $\gamma_e<$\textit{qed\_g\_cutoff}, particles are pushed by the LL pusher and if $\gamma_e>$\textit{qed\_g\_cutoff}, quantum Monte-Carlo approach is used.

\begin{figure}[h]
\centering
	\includegraphics[width=.99\textwidth]{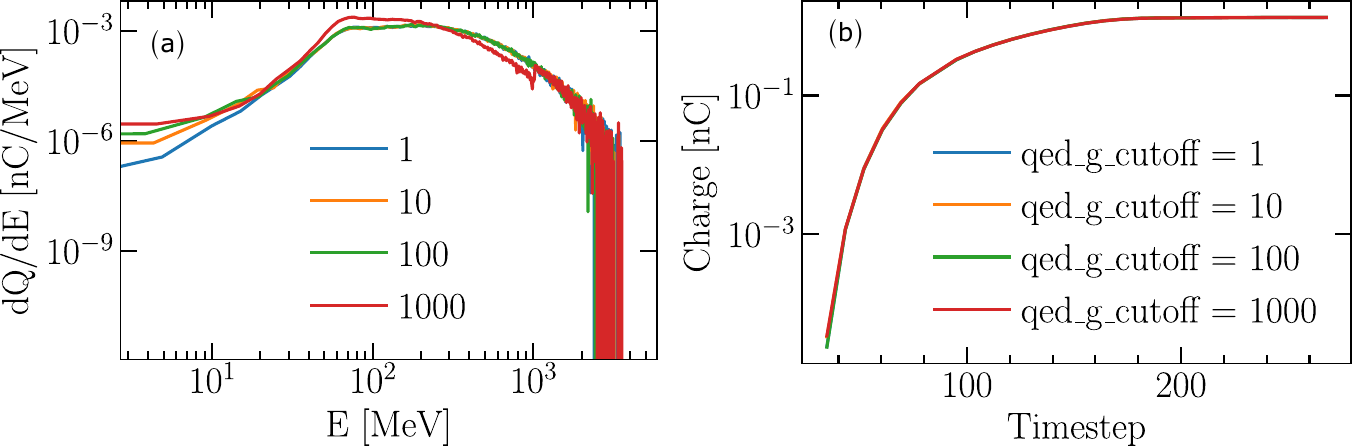}
	\caption{Study of numerical effects when varying the \textit{qed\_g\_cutoff}  parameter. (a) Spectra of positrons created. The discontinuity in the red curve at 1 GeV (\textit{qed\_g\_cutoff}$=1$ GeV for this case) is caused by overestimating the energy loss by the classical LL formula for particles below 1 GeV. (b) The number of generated positrons. Even though the shape of the spectrum can be affected by a high \textit{qed\_g\_cutoff}, in our particular case it had no impact on the number of generated pairs.  }
    \label{fig:qed_app} 
\end{figure}

In simulations where we study the single laser scheme to probe QED effects after the reflection from overdense foil, we use the \textit{qed\_g\_cutoff}$=800$, which means that particles with energies below 400 MeV are treated using LL radiation reaction and above 400 MeV are treated using Monte Carlo routine. For the DLA runs, it is sufficient to treat all electrons using classical LL approach. However, after the reflection, the value of $\chi_e$ rapidly changes due to the interaction with the counter-propagating laser. As a result of this, in Fig. \ref{fig:scan_position} (b) the discontinuity in the positron spectrum is shown. The reason is, that it is known that at values of $\chi_e$ at a fraction of unity, classical radiation reaction overestimates the energy damping compared to the correct QED treatment. However, this does not influence the number of generated positrons, since only electrons at energies at order of GeV contribute significantly to the pair creation. 

To demonstrate that keeping the parameter \textit{qed\_g\_cutoff}$=800$ during the whole simulation does not influence the number of generated pairs, we initialized a benchmark simulation where the threshold parameter was varied to values 2000, 200, 20 and 1. Simulation corresponds to the case of a 10 PW externally-guided laser pulse being reflected from the foil located at the position 4.2 mm. Fig. \ref{fig:qed_app} (a) shows that if the threshold energy is 1000 MeV, the discontinuity is present as expected. However, spectra of positrons for lower values of the parameter are almost identical, with only a small differences at low energy part. The panel (b) shows the number of positrons generated at simulations with different values of the parameter used and no difference is seen in the number. This agrees with our interpretation that overestimation of the LL radiation reaction that influences the shape of the spectrum has no influence on the number of generated positrons, because only electrons with energies $E>1$ GeV contribute significantly to the pair creation.

\bibliography{bibliography}

\end{document}